\documentstyle[epsfig]{aipproc}

\begin{document}

\newcommand{\beq}{\begin{equation}}
\newcommand{\eeq}{\end{equation}}
\newcommand{\sg}{\sigma}
\newcommand{\nc}{\newcommand}
\nc{\lapp}{\mbox{\raisebox{-.6ex}{$\,\stackrel{\textstyle <}{\sim}\,$}}}
\nc{\gapp}{\mbox{\raisebox{-.6ex}{$\,\stackrel{\textstyle >}{\sim}\,$}}}

\renewcommand{\thefootnote}{\fnsymbol{footnote}}

\protect\title{Searching for a $W'$ at the\\ 
Next-Linear-Collider using Single Photons\footnote%
{Supported in part by the Natural Sciences and Engineering 
Research Council of Canada.}}

\author{Stephen Godfrey, Pat Kalyniak and Basim Kamal}

\address{Ottawa-Carleton Institute for Physics \\
Department of Physics, Carleton University, Ottawa, ON K1S 5B6, Canada}

\author{Arnd Leike}
\address{LMU, Sektion Physik, Theresienstr.\ 37 , D-80333 M\"{u}nchen, Germany}
\maketitle

\begin{abstract}
We examine the sensitivity of the process $e^+e^- \rightarrow \nu\bar{\nu}
+ \gamma$ to additional $W$-like bosons which arise in various models.
The process is found to be sensitive to $W'$ masses up to several TeV.
\end{abstract}

\section*{INTRODUCTION}

There have been many studies of processes sensitive to additional $Z$-like
bosons ($Z'$s)  but comparatively few studies pertinent to 
$W'$s (see for instance \cite{BR,Hew}), especially at $e^+e^-$ colliders. Why?
Firstly, there are fewer models which predict $W'$s. 
Secondly, at LEP II energies,
the $W'$ signal is rather weak. Hence direct
searches have been limited to hadron colliders and the $W'$-quark
couplings are poorly constrained in some models.

In this contribution we present preliminary results of an
investigation of the sensitivity of the 
process  $e^+e^- \rightarrow \nu\bar{\nu}+ \gamma$ to
$W'$ bosons in various models. Our results are obtained by measuring the
deviation from the standard model expectation. Interesting 
discovery limits are
obtained for center-of-mass $e^+e^-$ energy of 500 GeV or higher
-- {\em Next-Linear-Collider}\ (NLC) energies.

Direct searches have been performed and indirect bounds
have been obtained for $W'$s in a few models, the details of which
are given later. Bounds for the Left-Right Symmetric Model (LRM)
and Sequential Standard Model (SSM) 
can be found in \cite{pdg}. They are obtained from the non-observation
of direct production
of $W'$s at the Tevatron and from indirect
$\mu$-decay constraints. For the LRM (with equal left- and
right-handed couplings) CDF obtains
$M_{W'} \gapp 650$ GeV and for the SSM, D0 finds
$M_{W'} \gapp 720$ GeV. From $\mu$-decay, the LRM $W'$ is constrained
to $M_{W'} \gapp 550$ GeV \cite{Barenboim}. 
A naive leading order analysis for the SSM
yields a bound of between 900 GeV and 1 TeV. One expects a somewhat higher
bound than for the LRM since, in $\mu$-decay, there will be a 
$W$-$W'$ interference term. The major limitation in this method is the
uncertainty in the $W$ mass.

The LHC will have a discovery reach in the TeV range \cite{LHC}. 
The search is analogous to
that done at the Tevatron, except for the higher energy and luminosity.
On the down side, one has $pp$ instead of $p\bar{p}$, which means no
valence-valence contribution in the large Feynman-$x$ region. For the
LRM, the magnitude of the effect will depend also on the $W'$-quark couplings
which are unknown. Therefore it is hard to make predictions
a-priori concerning discovery limits.
The NLC search nicely sidesteps the above problem as no
$W'$-quark couplings enter. Other LHC disadvantages include a lack of
initial state quark polarizability, parton distribution dependence and
large QCD corrections. The latter problems will affect the
ability to pin down the $W'$ couplings. Hence, the complimentary 
nature of the cleaner NLC measurement
is obvious despite the LHC's high energy
reach.

\section*{BASIC PROCESS}

\begin{figure}[t!] 
\centerline{\epsfig{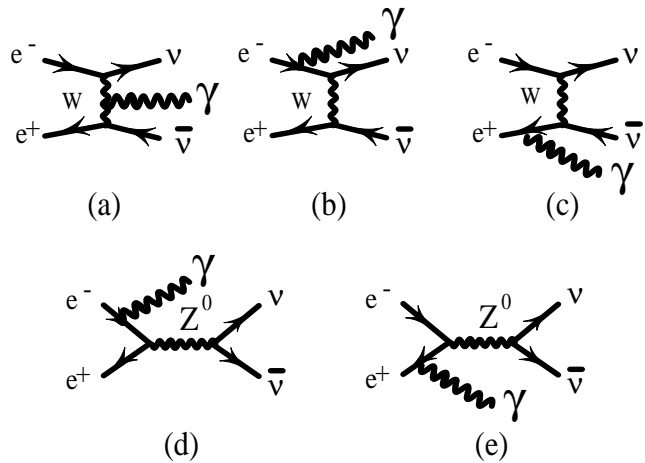}}
\vspace{10pt}
\caption{The Feynman diagrams contributing to the SM process.}
\label{feyn}
\end{figure}

The basic process under consideration is:
\begin{equation}
e^-(p_1) + e^+(p_2) \rightarrow \gamma(p_3) + [\nu(p_4) + \bar{\nu}(p_5)],
\end{equation}
where the square brackets indicate that since the neutrinos are not 
observed, we effectively only have single photon production. The diagrams
representing the leading Standard Model (SM) contribution are shown in
Fig.\ 1. The $W'$($Z'$) contributions are obtained by replacing
$W\rightarrow W'$, $Z\rightarrow Z'$ in the SM diagrams. Then one must
include all interferences between SM and beyond-SM diagrams in the 
squared amplitude.

The resulting squared amplitude is quite short, including spin dependence
which comes out automatically when expressing the result in terms of 
the left- and right-handed couplings. The result is quite general and
includes an arbitrary number of $W'$s and $Z'$s.

\section*{MODELS}

We have considered three models having $W'$s which contribute to our
process and are briefly described below.

{\em Sequential Standard Model:}\ This is the simplest $W'$-containing
extension of the SM, although not well motivated by theory. One has an extra
$W'$ which is heavier than the SM one, but which has identical couplings.

{\em Left-Right Symmetric Model:}\ In this model \cite{LRM}, the symmetry
$SU(3)_c\times SU(2)_L\times SU(2)_R \times U(1)_{B-L}$ is obeyed, 
giving rise to a
 $W'$ and a $Z'$. The $W'$ is purely right-handed; we do not
consider mixing between the SM and beyond-SM bosons. The pure SM couplings
remain unchanged and we take the new right-handed neutrinos to be 
massless. In principle they could be very heavy as well, but this would
lead to decoupling of the $W'$ from our process
and we would be effectively left with a
$Z'$ model, which is not the principal interest of this study.

Two parameters arise; $\rho$ and $\kappa$. For symmetry breaking 
via Higgs
doublets (triplets) $\rho=1$ (2). $\kappa$ is defined by $\kappa=g_R/g_L$
and thus measures the relative strength of the $W'l\nu_l$ and
$Wl\nu_l$ couplings. It lies in the range \cite{Hew,kappa}
\begin{equation}
\label{kappa_range}
 0.55 \lapp \kappa \lapp 2 \,\,.
\end{equation}
More specifically, we have the coupling
\beq
W'l\nu = i \frac{g\kappa}{\sqrt{2}} \gamma^\mu \frac{1+\gamma_5}{2},
\eeq
suggesting that larger values of $\kappa$ will lead to larger deviations
from the SM. In addition, we have the relation
\beq
M_{Z'}^2 = \frac{\rho \kappa^2}{\kappa^2-\tan^2\theta_W} M_{W'}^2,
\eeq
so that $\rho=1$ leads to a lighter $Z'$ mass for fixed $M_{W'}$, 
which should yield
a bigger effect versus $\rho=2$.

{\em Un-Unified Model (UUM):}\ The UUM \cite{UUM} obeys the symmetry 
$SU(2)_q\times SU(2)_l \times U(1)_Y$, again leading to a
$W'$ and a $Z'$. Both new bosons are left-handed and generally taken
to be approximately equal in mass. There are two parameters:
a mixing angle $\phi$, which represents a mixing between the charged
bosons of the two SU(2) symmetries,
and $x=(u/v)^2$, where $u$ and $v$ are the
VEV's of the two scalar multiplets of the model. 
The relation $M_{Z'}\simeq M_{W'}$ follows in
the limit $x/\sin^2\phi \gg 1$ and the parameter $x$ may
be replaced by $M_{W'}$, so that only $\phi$ enters as a parameter for
determining mass discovery limits. The leptonic couplings may be inferred from
the Lagrangian
\beq
{\cal L}_{\rm lept} =  - \frac{g \sin\phi}{2 \cos\phi} [ 
\sqrt{2} \,\, \overline{\psi}_{\nu_l} \gamma_\mu \psi_{l,L}  W_2^{+,\mu}
+ ( \overline{\psi}_{\nu_l} \gamma_\mu \psi_{\nu_l,L}
   -\overline{\psi}_{l} \gamma_\mu \psi_{l,L}) Z_2^\mu
]
,
\eeq
where $\psi_L = \frac{1}{2}(1-\gamma_5)\psi$. The existing constraint
on $\phi$ is \cite{BR}
\beq
\label{phi_range}
0.24 \lapp \sin\phi \lapp 0.99 \,\, .
\eeq

\begin{figure}[t!]
\vspace{1cm}
\includegraphics{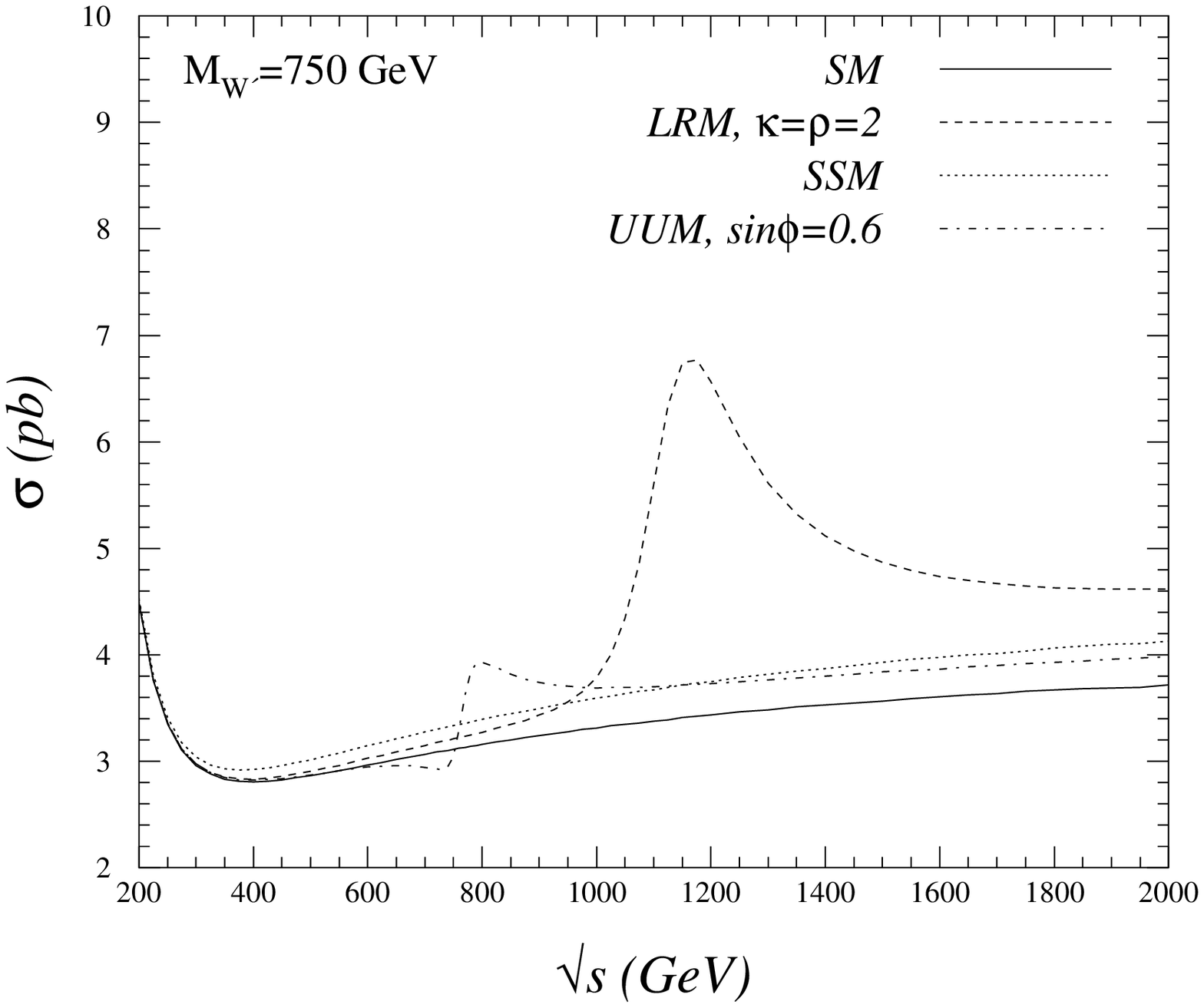}
\includegraphics{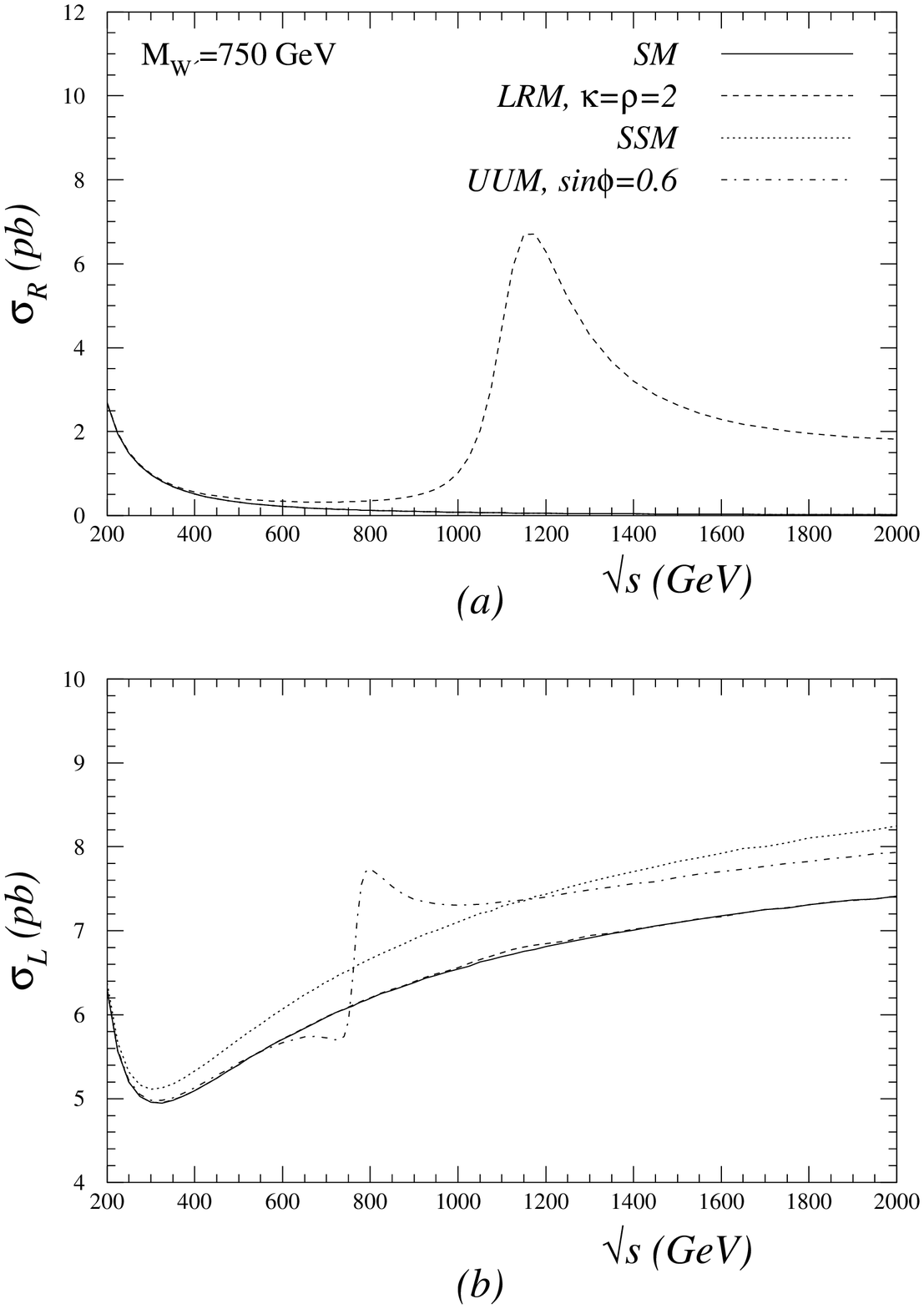}
\vspace{2.8in}
\vspace{1cm}
\begin{minipage}[b]{2.8in}
\caption{Unpolarized total cross section versus center-of-mass
energy in the SM, LRM, SSM and UUM.}
\label{sguvs}
\end{minipage}
\hfill
\begin{minipage}[b]{2.8in}
\protect\caption{As Fig.\ 2, except for (a) right-handed; (b) 
left-handed $e^-$ beam.}
\label{sglvrs}
\end{minipage}
\end{figure}

\section*{CROSS SECTIONS}

As inputs, we take $M_W = 80.33$ GeV, $M_Z = 91.187$ GeV, 
$\sin^2\theta_W = 0.23124$, $\alpha=1/128$, $\Gamma_Z=2.49$ GeV.
Let $E_\gamma$, $\theta_\gamma$ denote the photon's energy and angle
in the $e^+e^-$ center-of-mass, respectively.
No binning or transverse momentum cuts have
been explicitly introduced at this point. However, we have restricted
the range of $E_\gamma$, $\theta_\gamma$ as follows:
\beq
E_\gamma > 10\,\, {\rm GeV}, \,\,\,\,\,\,\,\,
10^o < \theta_\gamma < 170^o\,\, ,
\eeq
so that the photon may be detected cleanly. As well, the angular cut
eliminates the collinear singularity arising when the photon is 
emitted parallel to the beam.

\begin{figure}[t!]
\vspace{1.5cm}
\includegraphics{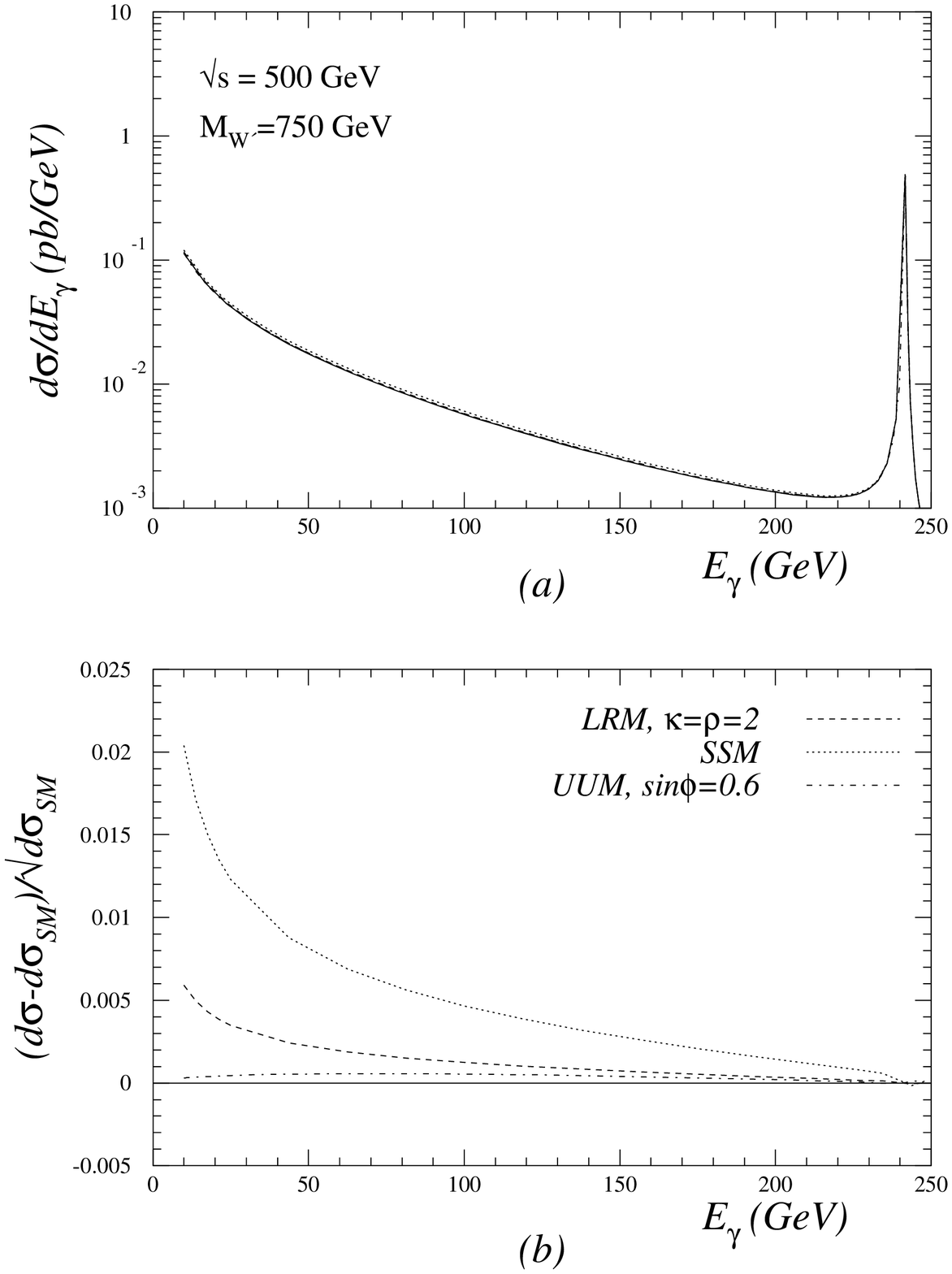}
\includegraphics{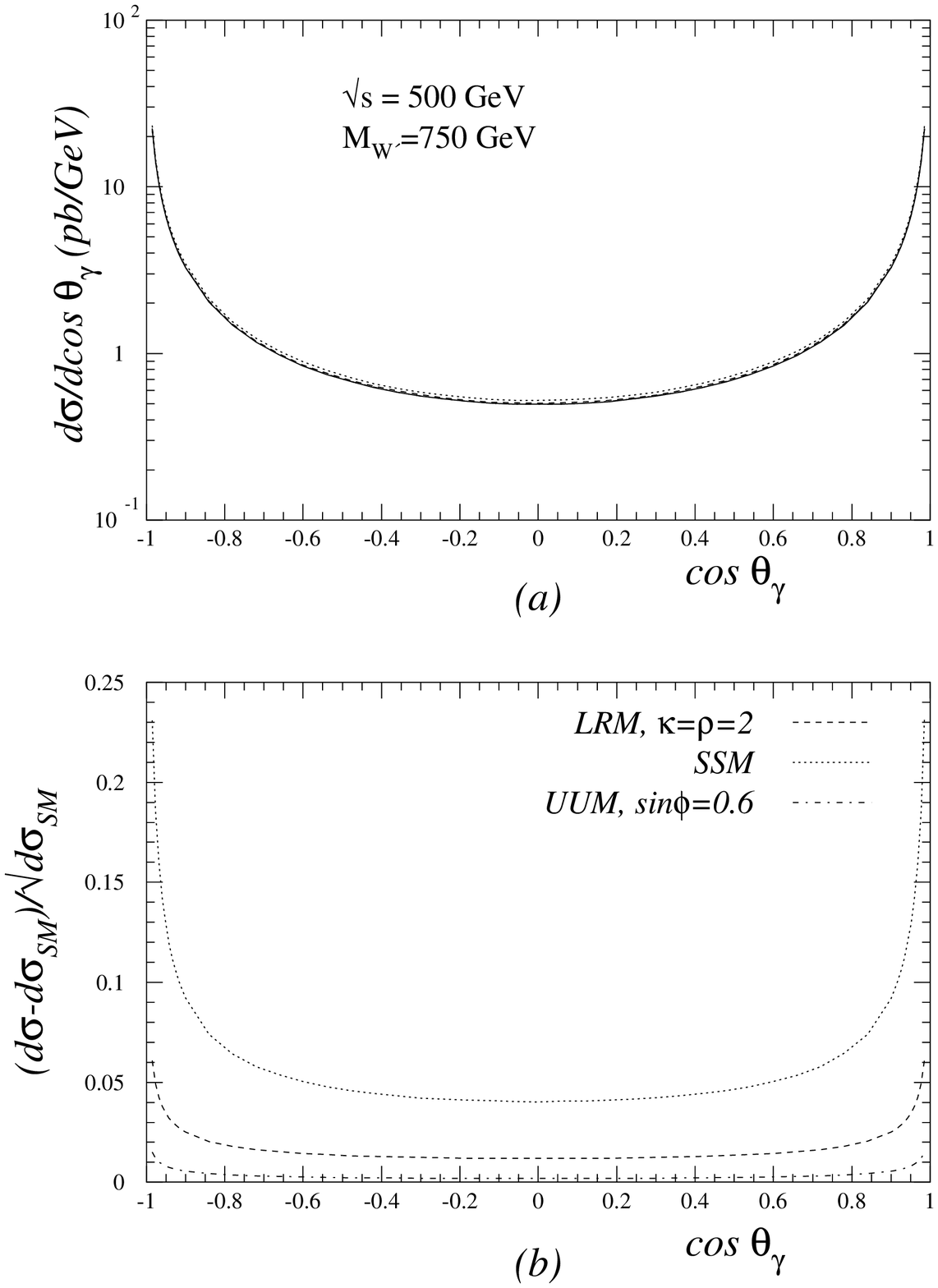}
\vspace{2.8in}
\vspace{1cm}
\begin{minipage}[b]{2.8in}
\caption{(a) Cross section and (b) relative statistical significance of the
deviation from the SM versus photon energy.}
\label{sguve5}
\end{minipage}
\hfill
\begin{minipage}[b]{2.8in}
\caption{(a) Cross section and (b) relative statistical significance of the
deviation from the SM versus $\cos\theta_\gamma$.}
\label{sguvz5}
\end{minipage}
\end{figure}

Figure 2 shows the total unpolarized cross section
versus center-of-mass energy for the SM, LRM (taking
$\kappa=\rho=2$), SSM and UUM (taking a representative value of
$\sin\phi = 0.6$). Throughout,
we use $M_{W'}$ = 750 GeV. Figure 3 shows the same, except for pure
right- and left-handed $e^-$ beams. The peaks are due to the $Z'$s in the
LRM and UUM. As expected, the LRM gives a large effect when the
$e^-$ is right-handed while the SSM and UUM give a larger effect for
a left-handed $e^-$.
The fact that the SM right-handed cross section goes to zero for large
$\sqrt{s}$ indicates $W$ ($t$-channel) dominance well above the
$Z$ pole.

In Figure 4(a), we plot the differential cross section with respect 
to $E_\gamma$ for $\sqrt{s}$ of 500 GeV.
The peak in the photon distributions is due to the radiative
return to the $Z$ resonance.
At higher energies, there are additional peaks in the $E_\gamma$ spectrum 
due to $Z'$s. We see that most of the 
contribution comes from the lower $E_\gamma$ range. This must be weighted
with the deviation from the SM in order to gauge the relative
statistical significance of the various energy regions. This is done
in Figure 4(b) where $(d\sg/dE_\gamma-d\sg_{SM}/dE_\gamma)/
\sqrt{d\sg_{SM}/dE_\gamma}$ is plotted as a function of
$E_\gamma$. Indeed, we see
that one benefits little from the region $E_\gamma$ above 
$\sim 200$ GeV in all
models. 

In Figure 5 we plot the the differential cross section with respect 
to $\cos\theta_\gamma$ and the corresponding relative
statistical significance.
We see that both
the cross section and relative statistical significance are peaked in
the forward/backward direction and the distributions are very nearly
symmetric in $\cos\theta_\gamma$. In Figures 4(b) and 5(b), the overall
normalization is unimportant. Experimentally, some binning scheme
will be adopted and each bin will carry a weight proportional to the
beam luminosity.

\section*{NLC $W'$ mass discovery Limits}

\begin{figure}[t!] 
\vspace{1.5cm}
\includegraphics{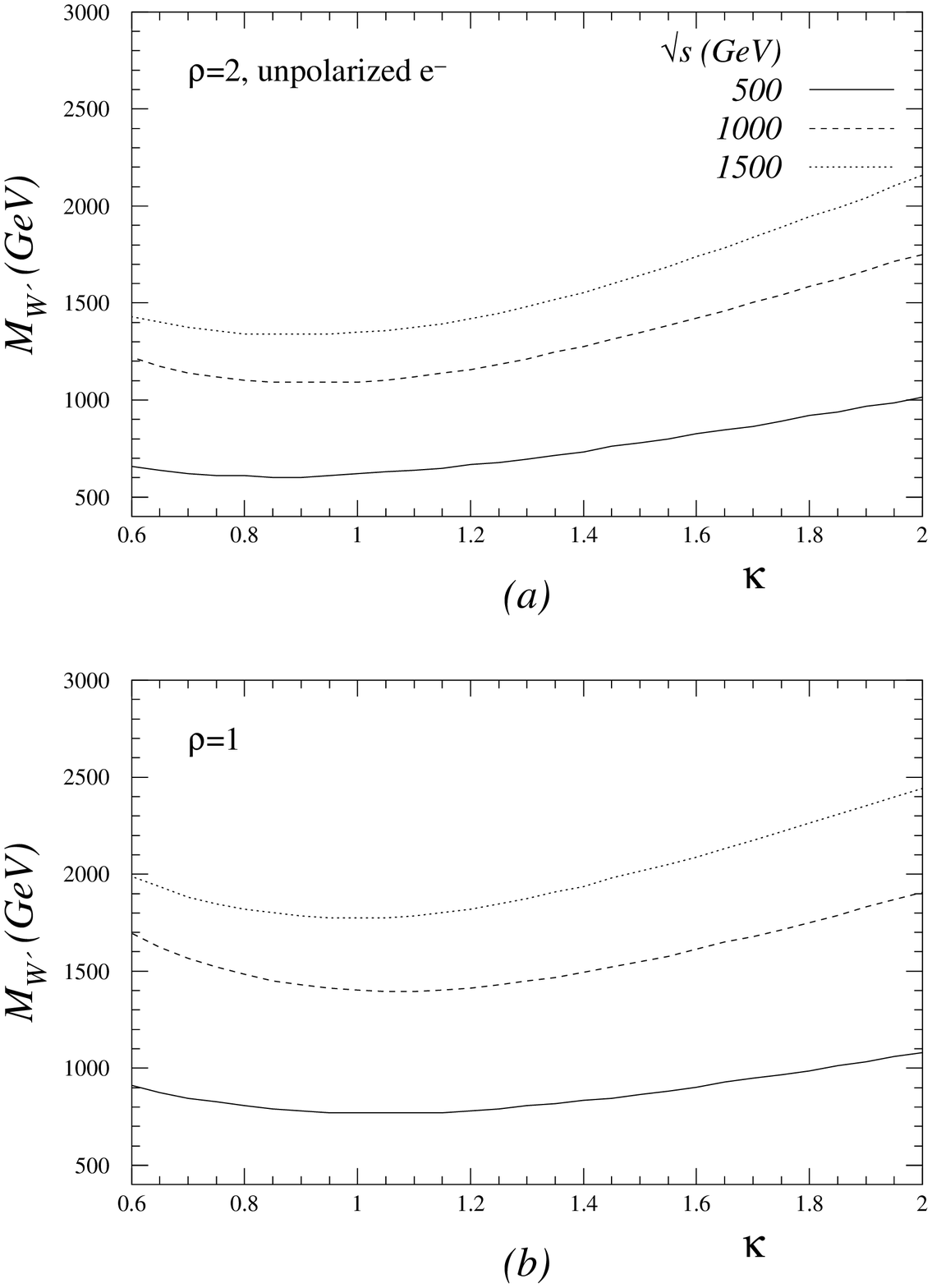}
\includegraphics{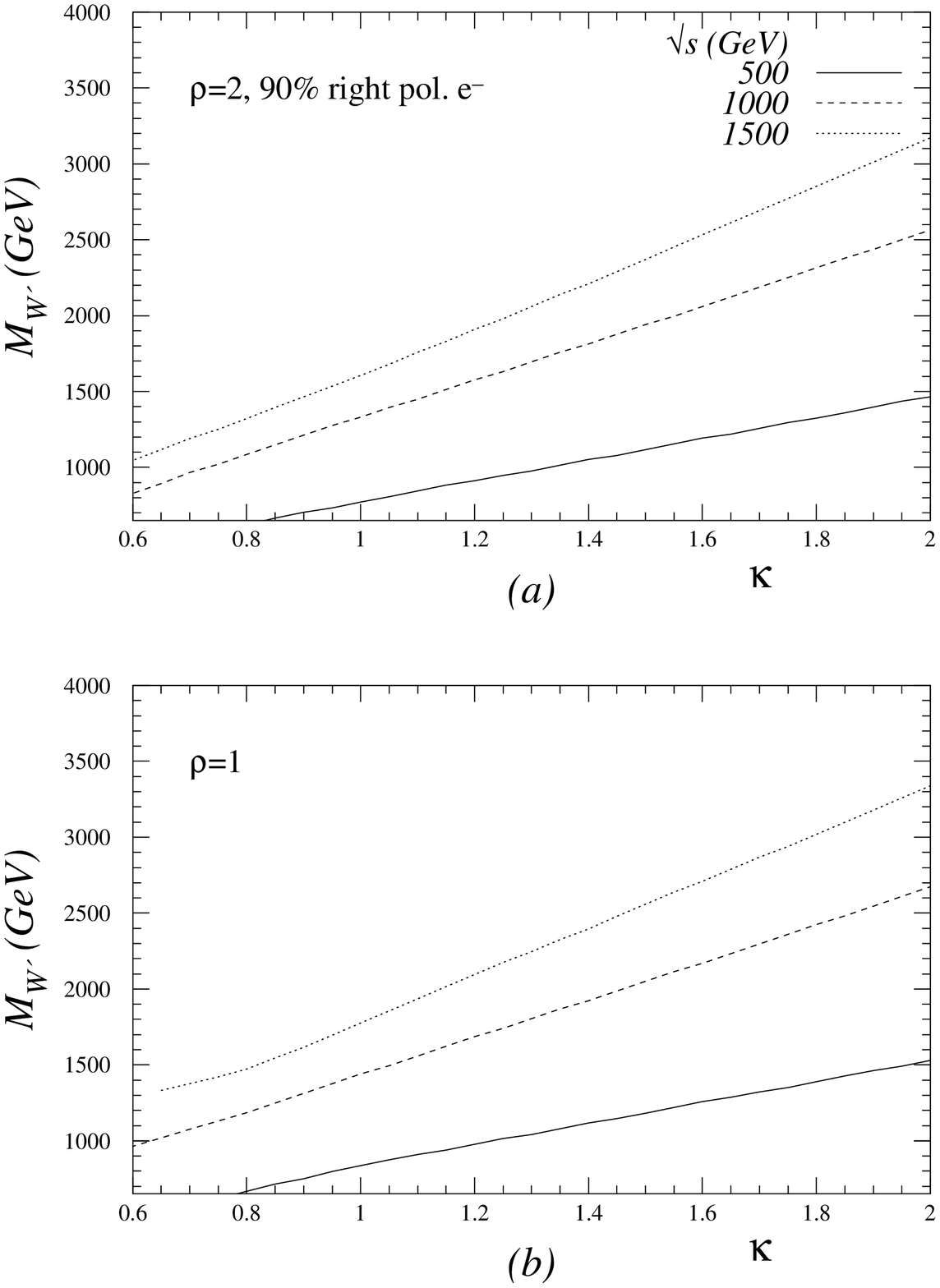}
\vspace{2.8in}
\vspace{1cm}
\begin{minipage}[b]{2.8in}
\caption{$W'$ 95\% C.L. discovery limits versus \protect$\kappa$ in the LRM 
obtained with unpolarized beam for (a) $\rho=2$; (b) $\rho=1$.
}
\label{limuvkap}
\end{minipage}
\hfill
\begin{minipage}[b]{2.8in}
\caption{As Fig.\ 6, except
with 90 percent right-polarized electron beam.}
\label{limrvapp9}
\end{minipage}
\end{figure}

For $\sqrt{s}$ of 500 GeV, we take an integrated luminosity 
of $50 fb^{-1}$ and for $\sqrt{s}$ of 1 TeV and 1.5 TeV we take
$200 fb^{-1}$. For the polarized limits we take half
the above luminosities assuming equal running in both polarization
states (of the $e^-$ beam). We assume 90\% $e^-$ polarization 
unless otherwise stated. 

In obtaining limits, we have imposed the
additional cut:
\beq
E_\gamma < E_{\gamma, {\rm max}} \,\, ,
\eeq
in order to cut out the  high energy
events, especially those near the $Z$ pole,
which are insensitive to 
$W'$s and $Z'$s.
It was found that $E_{\gamma, {\rm max}}$ of 200, 350 and 500 GeV
for $\sqrt{s}$ of $500$ GeV, $1$ TeV and $1.5$ TeV, respectively,
lead to the best limits in general, although the limits 
were not very sensitive to moderate variations in $E_{\gamma, {\rm max}}$.
The limits are given at 95\% confidence level and are 
calculated for all three energies.

Figure 6 presents the $W'$ mass
discovery limits obtainable with an unpolarized beam
for the LRM, plotted versus $\kappa$ for $\rho=1,2$. Depending on 
$\sqrt{s}$, $\rho$ and $\kappa$, they range from 600 GeV to 2.5 TeV.
The predicted dependence on $\kappa$ and $\rho$ is generally observed,
except at low $\kappa$ where we notice a moderate {\em increase} in the
limits, even though the $W'$ couplings have weakened. We attribute this
effect to the $Z'$, whose couplings are enhanced (but its mass increased)
in the low $\kappa$ region. This is evidenced by the appreciable 
improvement in the bounds for low $\kappa$ and $\rho=1$. Figure 7 demonstrates
the improvement in bounds in the moderate to large $\kappa$ region
obtained when a (90\%) polarized right-handed $e^-$ beam is used. The
beam polarization picks out the LRM $W'$ and suppresses the SM $W$.
Increase of the polarization can lead to even higher
limits as shown in the rightmost column of Table 1, where limits are
tabulated for all three models.

\begin{table}
\caption{$W'$ 95\% C.L.\ discovery 
limits obtained in the SSM, LRM ($\kappa=\rho=2$)
and the UUM ($\sin\phi=0.6$), assuming 90 percent $e^-$ polarization 
(unless otherwise stated) and
using $1/2$ the unpolarized luminosity for the left and right cases.}
\label{tablimits}
\begin{tabular}{llllll}
   \parbox{1cm}{\protect $\sqrt{s}$  (GeV)} & Model &
 \parbox{25mm}{\raggedright Unpolarized \protect $e^-$  Limit (TeV)}
& \parbox{22mm}{\raggedright Left Pol.\ \protect $e^-$  Limit (TeV)}
& \parbox{22mm}{\raggedright Right Pol.\ \protect $e^-$  Limit (TeV)}
& \parbox{25mm}{\raggedright 100\% L/R Pol.\  Limit (TeV)}
\\
\tableline
500 & SSM & 2.45 & 2.45 & 1.15  & 2.45\\
(50 fb$^{-1}$) & LRM & 1.0 & $<$0.75 & 1.45 & 2.05 \\
 & UUM & 0.65 & 0.65 & 0.55 & 0.65 \\
 & & & & & \\
1000 & SSM & 4.55 & 4.5 & 2.15 & 4.55 \\
(200 fb$^{-1}$)  & LRM & 1.75 & $<$1 & 2.55 & 4.5 \\
 & UUM & 1.3 & 1.3 & 1.15 & 1.3 \\
 & & & & & \\
1500 & SSM & 5.2 & 5.15 & 2.45 & 5.2 \\
(200 fb$^{-1}$)  & LRM & 2.15 & $<$1.25 & 3.2 & 6.2 \\
 & UUM & 1.85 & 1.85 & 1.65 & 1.85 \\
\end{tabular}
\end{table}

The highest limits are obtained for the
SSM in most scenarios, except when $\sin\phi$ is large, as indicated in
Figure 8 which shows the limits in the UUM versus $\sin\phi$. We observe
a turn-on in sensitivity for $\sin\phi\gapp 0.62$ at
$\sqrt{s} = 500$ GeV and 1 TeV, while for $\sqrt{s} = 1.5$ TeV this
occurs for $\sin\phi\gapp 0.73$. The interference term may play
a role in this behaviour.
From another perspective,  for fixed $\sin\phi$,
one may observe sudden changes in sensitivity as $\sqrt{s}$ is varied
as can be seen from the changing of the sign of the effect on the cross
section in Figs 2,3. The result is that for $0.62 \lapp \sin\phi
\lapp 0.72$, we obtain better limits at $\sqrt{s} = 1$ TeV than we
do at  $\sqrt{s} = 1.5$ TeV.

\begin{figure}[t!] 
\includegraphics{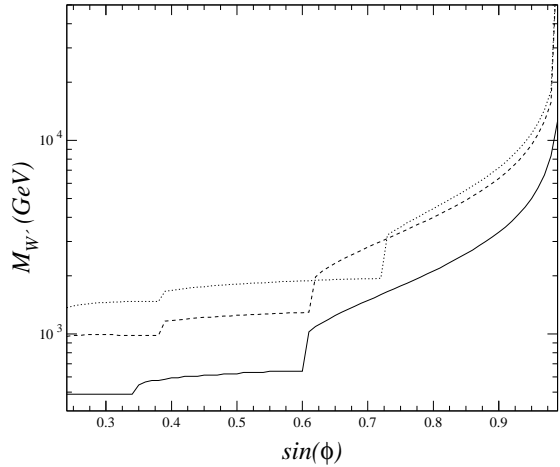}
\vspace{2.5in}
\caption{ 95\% C.L. discovery limits versus 
$\sin(\phi)$ in the UUM; lines as in Fig.\ 6.}
\label{limvphi}
\end{figure}

For both the UUM and the SSM, where the $W'$s are left-handed, there is
little benefit from polarization. The reason is that all the effect
comes from the left-handed $e^-$ initial state, which also dominates the
unpolarized cross section. After folding in the luminosity decrease
associated with running in a particular polarization state, all 
benefits are lost.

\section*{SUMMARY AND OUTLOOK}

The usefulness of the process $e^+e^- \rightarrow \nu\bar{\nu}
+ \gamma$ in searching for $W'$s has been demonstrated and should
be complimentary to direct searches at the LHC.
The results of our study will be extended to include
cuts on the transverse momentum of the photon to reduce backgrounds
(primarily from radiative Bhabha scattering with undetected $e^+e^-$)
and to examine the effect of binning. All remaining backgrounds will
have to be included in the final analysis of the data and are currently
under investigation, but are not expected to significantly affect our limits. 
Other models are also under consideration.


\begin{references}
\bibitem{BR}Barger, V., and Rizzo, 
T., {\it Phys.\ Rev.}\ {\bf
D41}, 946 (1990). 
\bibitem{Hew}Hewett, J., SLAC-PUB-7441, June 1996, hep-ph/9704292.
\bibitem{pdg}Particle Data Group (Caso, C., et al.), {\it Eur. Phys. J.}\ 
{\bf C3}, 1 (1998).
\bibitem{Barenboim}Barenboim, G., Bernab\'{e}u, J., Prades, J.,
and Raidal, M., {\it Phys.\ Rev.} {\bf D55}, 4213 (1997).
\bibitem{LHC}For a review, see: Cveti\u{c}, M., and Godfrey, S., 
``Discovery and Identification
of Extra Gauge Bosons,'' in {\it Electro-weak Symmetry Breaking and Beyond
the Standard Model}, eds. Barklow, T., et al., World Scientific, 1995,
hep-ph/9504216.
\bibitem{LRM}Mohapatra, R.N., {\it Unification and Supersymmetry},
New York: Springer, 1986, and original references therein.
\bibitem{kappa}Parida, M., and 
Raychaudhuri, A., {\it Phys.\ Rev.}\ {\bf D26}, 2364 (1982);
Chang, D., Mohapatra, R., and Parida, M.,
{\it Phys.\ Rev.}\ {\bf D30}, 1052 (1984).
\bibitem{UUM}Georgi, H., Jenkins, E.E., and Simmons, E.H., {\it Phys.
Rev. Lett.}\ {\bf 62}, 2789 (1989); {\bf 63}, 1540(E) (1989).
\end{references}
\end{document}